\documentclass[prb,twocolumn,aps,showpacs]{revtex4}
\usepackage{graphicx}
\usepackage[activeacute,english]{babel}

\usepackage{amsmath}
\usepackage{amsfonts}

\begin{document}

\title{Crossed-ratchet effects and domain wall geometrical pinning }

\author{V.I. Marconi}
\affiliation{Facultad de Matem\'atica,
Astronom{\'i}a y F{\'i}sica,   Universidad Nacional de C\'ordoba
  and  IFEG-CONICET, X5000HUA C\'ordoba, Argentina.}

\author{A.B. Kolton}
\affiliation{CONICET, Centro At\'omico Bariloche, 8400 S.C. de Bariloche, Argentina.}

\author{J.A. Capit\'an}
\affiliation{Dept. Matem\'aticas and {\em GISC}, Universidad
Carlos III de Madrid, 28911 Legan\'es, Spain.}

\author{J.A. Cuesta}
\affiliation{Dept. Matem\'aticas and {\em GISC}, Universidad
Carlos III de Madrid, 28911 Legan\'es, Spain.}

\author{A. P\'erez-Junquera}
\affiliation{Dept. F{\'i}sica, Universidad de Oviedo-CINN, 33007
Oviedo, Spain.}

\author{M. V\'elez}
\affiliation{Dept. F{\'i}sica, Universidad de Oviedo-CINN, 33007
Oviedo, Spain.}

\author{J.I. Mart{\'i}n}
\affiliation{Dept. F{\'i}sica, Universidad de Oviedo-CINN, 33007
Oviedo, Spain.}

\author{J.M.R. Parrondo}
\affiliation{Dept. F{\'i}sica At\'omica, Molecular y Nuclear and
{\em GISC}, Universidad Complutense de Madrid, 28040 Madrid,
Spain.}

\
\date{\today}

\begin{abstract}

The motion of a domain wall in a two dimensional medium is studied
taking into account the internal elastic degrees of freedom of the
wall and geometrical pinning produced both by holes and sample
boundaries. This study is used to analyze the geometrical
conditions needed for optimizing crossed ratchet effects in
periodic rectangular arrays of asymmetric holes, recently observed
experimentally in patterned ferromagnetic films. Geometrical
calculations and numerical simulations have been used to obtain
the anisotropic critical fields for depinning flat and kinked
walls in rectangular arrays of triangles. The aim is to show with a generic elastic model for interfaces how to build a rectifier able to display crossed ratchet effects or effective potential landscapes for controlling the motion of interfaces or invasion fronts.
\end{abstract}


\maketitle

\section{Introduction}

The dynamics of elastic interfaces is responsible for a wide
variety of physical phenomena in very different experimental
systems. Prominent examples are the propagation of reaction fronts
or surface growth in material science ~\cite{barabasi_book}, cell motility
and membrane dynamics in biology \cite{biology}, domain walls in
ferromagnetic \cite{prl,lemerle_domainwall_creep,krusin_pinning_wall_magnet,repain_avalanches_magnetic,caysol_minibridge_domainwall,Metaxas}
 or ferroelectric films \cite{tybell_ferro_creep,paruch_2.5,kleeman},
fluid invasion in porous media \cite{wilkinson_invasion}, contact lines of liquids menisci
\cite{moulinet_contact_line},
and crack propagation \cite{bouchaud_crack_propagation,alava_review_cracks}. In all these
cases, the presence of heterogeneities, which locally promote wandering,
compete with the elasticity of the interface, giving rise to complex
collective pinning effects. Understanding these effects is a
challenging problem relevant both for the basic and applied
viewpoint.

A particularly interesting case of interface pinning is the
``geometrical pinning'' induced by the presence of artificially
introduced holes or antidots~\cite{antidots,antidotsb,rodriguez},
or by a spatial modulation of the sample boundary conditions in
narrow samples
\cite{caysol_minibridge_domainwall,notches1,notches2,angelfish,allwood,notches3}. These kind
of boundaries can pin the interface by locally reducing its
extension, thus saving surface tension energy. For extended domain
walls this kind of pinning has been recently realized
experimentally and showed to be able to modify the magnetization
dynamics
\cite{antidots,rodriguez,caysol_minibridge_domainwall,notches1,notches2}
and to produce, in particular, interesting ratchet transport of
magnetic domain walls~\cite{prl,JPD}. Being mostly geometrical
(i.e. determined mostly by the shape and distribution of holes, or
by the geometry of the boundaries and not much on the specific
microscopic pinning interaction) this kind of pinning has the
advantage over other artificial pinning mechanisms that it can be
more easily tailored at a wide range of scales to control the wall
motion in various specific ways.

We have recently analyzed, specifically,  the pinning effect of
asymmetric holes on the propagation of domain walls in magnetic
films, finding that, under certain geometries and oscillating
external magnetic fields, the motion of flat and kinked walls is
rectified in opposite directions~\cite{prl}: the asymmetry between
forward/backward flat wall propagation results in a direct ratchet
effect whereas the asymmetry between upward/downward kink
propagation along a wall induces an inverted ratchet effect. This
striking sensitivity yields new strategies to control the motion of
the wall. The crossed rectification reported in Ref.~\cite{prl}
relies on the difference between the critical fields to depin the
wall in each direction, an it is also present in a generic model for
elastic interfaces: the $\phi^4$ model~\cite{Chaikin}. In this
paper we calculate the depinning field of a  generic $\phi^4$
interface in the presence of an array of triangular antidots both by
geometrical considerations and numerical simulations. We use this
simple model because  it is the minimum model that captures the
essential physics behind the crossed ratchet effects reported in \cite{prl}. In addition, our method can be widely used to design
interface rectifiers of elastic interfaces by using holes or
boundary conditions in an arbitrary geometry.

Our starting point is the overdamped $\phi^4$ model in the plane,
i.e., a scalar field $\phi(x,y;t)$ obeying the following evolution
equation:
\begin{equation}
\eta \partial_t \phi=c\nabla^2 \phi+\epsilon_0(\phi-\phi^3)+H
\label{phi4}
\end{equation}
where $c$ is the elastic stiffness of the order parameter,
$\epsilon_0$ is proportional to the local barrier separating two
minima of the local free energy, $H$ is an external field biasing
one of the two minima, and the friction coefficient $\eta$  sets the
microscopic time scale. The evolution equation (\ref{phi4}) derives
from the energy functional:
\begin{equation} E=\int
dxdy\left[U(\phi(x,y)) - H\phi(x,y)+\frac{c}{2}|\nabla\phi(x,y)|^2\right]
\label{energy0}
\end{equation}
with $U(\phi)=\epsilon_0(\phi^2-1)^2/4$.

 For $H=0$, and with the
appropriate boundary condition, say, $\phi(-L,y;t)=1$;
$\phi(L,y;t)=-1$, the stationary solution of Eq.~(\ref{phi4}) is
given by a domain on the left side of the plane with positive and
approximately homogenous field and a domain on the right side with
negative field, both separated by an interface of width
proportional to $\sqrt{c/\epsilon_0}$. When the field is switched
on to a positive (negative) value, the interface is pushed to the
right (left) to minimize the total energy. However, the interface
has also an elastic energy proportional to its length. Therefore,
if the geometry where the field is defined is such that the length
of the interface increases when moving to the left or right, then
the interface will be pinned until the field reaches a critical
value. Our goal is to provide an estimation of such depinning
field in a general geometry and to analyze the geometrical
conditions in which rectification effects appear in the elastic
interface propagation.

The organization of the paper is as follows. In
Sec.~\ref{sec:theory}, we develop a general theoretical framework to
address the problem of rectification of domain walls. We reduce the
field $\phi$ to an elastic wall and derive an analytical expression
for the local depinning field in two dimensional stripes with
arbitrarily shaped borders. In Sec.~\ref{sec:ratchet} we apply the
previous results to build a 2D array of triangular holes that
displays ratchet effects and give specific predictions for the
appearance of normal and crossed ratchet effects.
In Sec.~\ref{sec:conclusions} we summarize our results.

\section{Depinning fields in arbitrary geometry: general theory}
\label{sec:theory}

\subsection{From field equations to elastic walls}

Our aim is to calculate the depinning field of certain interfaces in
an arbitrary geometry. To simplify the task, we first need to reduce
the whole field equation (\ref{phi4}) to a parametric description of
the interface, in the spirit of the collective coordinate approach
widely used in one-dimensional models \cite{cca}.

In Appendix \ref{ap:line} we construct a solution of
Eq.~(\ref{phi4}) where two domains of positive and negative
magnetization are separated by a wall defined by the line
$(x(s),y(s))$. The solution reads:
\begin{equation} \phi(x,y)=\tanh\left[ \frac{ g(x,y)}{w}\right]
\label{eq:kink}
\end{equation}
where $g(x,y)$ is the distance of point $(x,y)$ to the backbone of the wall
$(x(s),y(s))$ and
\begin{equation}
w=\sqrt{\frac{2c}{\epsilon_0}}
\end{equation}
can be considered as its width. Eq.~(\ref{eq:kink}) is
well known as the field corresponding to a single wall in the one
dimensional $\phi^4$ model, and, as we show in Appendix \ref{ap:line}, can be extended to two dimensions if
the curvature radius of the wall $(x(s),y(s))$ is much larger than
its width $w$.

The energy of this solution, for small width $w$, can be
approximated by (see Appendix \ref{ap:line})
\begin{equation}
E=\sigma d-2HA
\label{energialineasinhuecos}
\end{equation}
where $d$ is the length of the wall $(x(s),y(s))$, $A$
the area of the positive domain (at one side of the
wall, and
\begin{eqnarray}
\sigma=\frac{\sqrt{8\epsilon_0c}}{3}
\end{eqnarray}
is the energy of the wall per unit of length. Consequently, the wall
behaves as an elastic line with a linear tension $\sigma$ and pushed
by a field $H$.

Our approximations are exact for infinitely narrow interfaces, $w
\to 0$, since we are reducing the field in the whole
plane to a single curve defining the center of the interface. For
thin interfaces the approximation is good enough, provided the width
of the wall remains approximately constant all along the curve and that the local curvature radius of
the line is smaller than the domain wall width $w$. In brief, these
conditions assure that the state of $\phi$ with a domain wall can be
well described exclusively by the transverse degrees of freedom of
an elastic interface.

\subsection{Interface pinning in a holed medium}
\label{sec:holepinning}

We will now consider a domain wall in a two
dimensional medium with holes or multiply connected space.
For the scalar field $\phi$ this amounts to solving Eq.~(\ref{phi4})
in a domain $\Omega-\bigtriangleup$, which includes all the two
dimensional space $\Omega$, except the possibly non-compact region
$\bigtriangleup$ occupied by the holes and outer space. In order to
model the absence of material in $\bigtriangleup$, we set free
(Neumann) boundary conditions for the order parameter,
$\partial_{\bf n} \phi|_{\partial \bigtriangleup} = 0$ at the hole
borders and sample boundaries ${\partial \bigtriangleup}$.

\begin{figure}
\begin{center}
\includegraphics[angle=0, width=8cm]{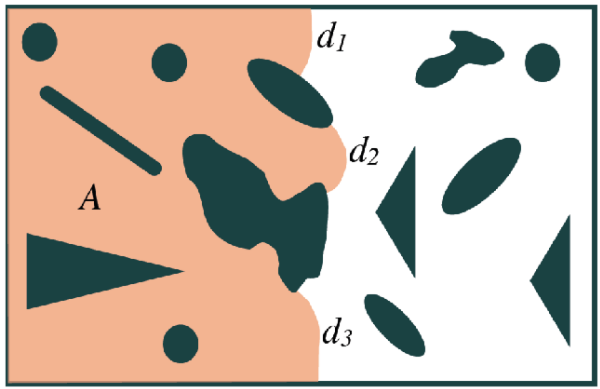}
\caption{(color online) Interface in a two dimensional medium with
free holes and sample boundaries. The gain of interface energy by
optimally intersecting the holes and sample boundaries produces
domain wall pinning.} \label{fig:nueva}
\end{center}
\end{figure}

Within the interface approximation described in the previous subsection,
interface pinning arises from the gain of line energy (reduction of
the total length of the interface) that is possible by optimally
intersecting the holes and sample boundaries (see
Fig.~\ref{fig:nueva}).  From Eq.~(\ref{energialineasinhuecos}) the
energy of the pinned domain wall then reads
\begin{eqnarray}
E = \sigma \sum_{i=0}^{N} d_i - 2 H A
\label{energialinea}
\end{eqnarray}
where $d_i = \int_{s_i}^{s_{i+1}} ds \sqrt{\dot{x}^2 + \dot{y}^2}$ is
the length of the interface segment
connecting the holes $i$ and $i+1$ (with $i=0$ and $i=N$ designating
the sample boundaries) and the area $A$
now excludes regions belonging to $\bigtriangleup$.

The free (Neumann) boundary conditions for the order parameter at the hole and sample
boundaries translate in the interface description in the orthogonality condition
\begin{eqnarray}
{\bf v}_i \cdot {\bf t}_i=0,  \quad \forall i
\label{perpcondition}
\end{eqnarray}
where ${\bf t}_i$ is the tangent vector of the boundary  and ${\bf
v}_i \equiv (\dot x(s_i),\dot y(s_i))$ the tangent vector of the
interface both at the intersection point $(x(s_i),y(s_i))$.

Metastable states of the interface are therefore local minima of
the energy (\ref{energialinea}) with segments satisfying the
orthogonality constraint (\ref{perpcondition}) at its ends. In the
following we discuss the geometry of these optimal segments, which
are the building blocks of our method.

\subsection{Equilibrium state of a wall}\label{sec:basic}

Our next step is to calculate the equilibrium profile of an
interface segment and its stability. The energy of an interface
segment is both a function of its shape and the location of its ends
or contact points. In order to find the possible metastable states of the
segment we need to minimize the energy given by
Eq.~(\ref{energialineasinhuecos}) with the constraint
(\ref{perpcondition}).

As shown in Appendix \ref{sec:basicap}, the solution of the corresponding Euler-Lagrange equation,
regardless of any boundary condition, is a circular arc of radius
\begin{equation}
r\equiv
\frac{\sigma}{2H}
\label{radius}
\end{equation}

\begin{figure}
\begin{center}
\includegraphics[angle=0, width=7cm]{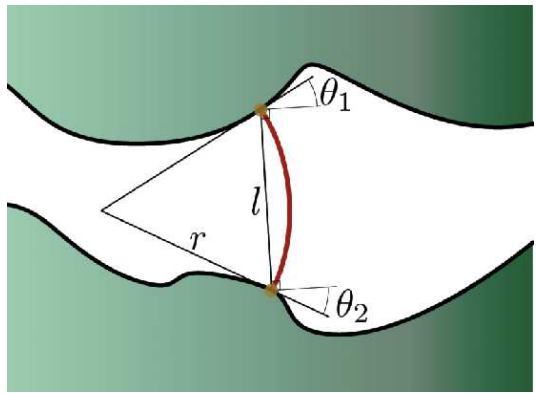}
\caption{(color online) The energy of an elastic wall (red online)
is minimized by an arc of radius $r=\sigma/(2H)$. An equilibrium
state is reached when the arc intersects the boundaries
orthogonally.} \label{fig:arc}
\end{center}
\end{figure}

Consider now a wall confined between two irregular boundaries, as
plotted in Fig.~\ref{fig:arc}. Let $l$ be the distance between the contact points, and $\theta_1+90^{\rm o}$ ($\theta_2+90^{\rm o}$) the angle formed
by the upper (lower) boundary and the line connecting the two contact points. The elastic wall minimizes its energy adopting the shape of
an arc of radius $r$ and it must be orthogonal to the boundaries at
the contact points. As illustrated in Fig.~\ref{fig:arc}, this implies that $\theta_1=\theta_2=\theta$ and
\begin{equation}
\sin\theta = \frac{l/2}{r} =\frac{ H l}{\sigma}
\label{anglecondition}
\end{equation}
where we have used the expression for the radius of the wall,
Eq.~(\ref{radius}).

\subsection{Local depinning fields for an anchored wall}\label{sec:depinning}

We can now proceed and calculate local depinning fields for narrow
domain walls bounded between two borders, which are central for
studying the ratchet effect. Given a metastable state of the
anchored domain wall the local depinning field is defined as the
maximum field it can support by deforming continuously as we
increase the field. Above this local depinning field the domain wall
escapes the local environment and slides until it is trapped again
in a new metastable state with a larger depinning field, if it
exists. Otherwise it continues sliding.

As an illustration, consider the particular case where the bottom
border is the $x$-axis and the top border is given by and arbitrary
smooth function $f(x)>0$. The wall, as we have seen above, is an arc
of radius $r=\sigma/(2H)$. Its center must lie in the $x$-axis, say
at $x_0$, since the wall is perpendicular to the $x$-axis at the
lower contact point. The upper contact point, $(x_1,f(x_1))$,
belongs to the arc, hence $(x_1-x_0)^2+f(x_1)^2=r^2$, and the
orthogonality condition implies:
\begin{equation*}
\frac{f(x_1)}{x_1-x_0}=f'(x_1)
\end{equation*}
The upper contact point is then given by the condition
\begin{equation}
r = \frac{f(x_1)}{f'(x_1)}\sqrt{1+f'(x_1)^2}
\label{solucion}
\end{equation}
Note that only the solutions with $f'(x_1) \ge 0$ must be taken if
$H \ge 0$. If such a solution $x_1$ exists for a given $r$ (i.e.,
for a given field $H$), the lower contact point is given by
$x_2=x_1$ if $f'(x_1) = 0$ and $x_2 = r + x_1 -f(x_1)/f'(x_1)$
otherwise.

For $H=0$ ($r\to \infty$) the only possible solutions are points
$x_1$ such that $f'(x_1)=0$. These solutions are straight vertical
segments joining the two borders at $x_1$. For concreteness let us
assume that $x_1=0$ for $H=0$ and that $f''(0)>0$, so the initial
state is metastable. In such a case if we
 quasistatically increase $H$ (decrease $r$)
from $x_1=0$ we can generate a continuum set of solutions $x_1(r)$
parametrized by the field. At some field $H_c=\sigma/2r_c$ it is
possible however to have a discontinuity in $x_1(r)$ due to the
absence of solutions beyond $H_c$. We can then define the
critical radius of the initial metastable state as
\begin{equation}
r_c = \min_{x_1} \left\{ \frac{f(x_1)}{f'(x_1)}\sqrt{1+f'(x_1)^2}\right\}_{f'(x_1)>0}
\label{solucionrc}
\end{equation}
where the condition $f'(x_1)>0$ assures that $r_c$ is positive, so
we obtain the forward depinning field. The depinning field is
therefore $H_c=\sigma/2r_c$, the upper contact point of the critical
arc is $x_c^{\rm up} \equiv x_1(r_c)$ and the lower contact $x_c^{\rm low} =
r + x_c -f(x_c)/f'(x_c)$.

\begin{figure}
\begin{center}
\includegraphics[angle=0, width=7cm]{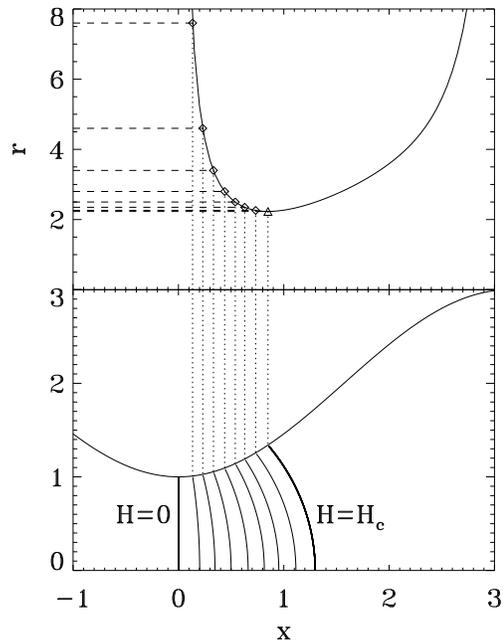}
\caption{Construction of the metastable elastic lines bounded
between $0$ and $f(x)=2-\cos(x)$, as a function of the external
field or arc radius. We depict the boundaries and the metastable
walls for different fields $H$ in the lower plot. In the upper
plot we represent the function $[f(x)/f'(x)]\sqrt{1+f'(x)^2}$. The
intersection of the arc radii $r=\sigma/(2H)$ with this function
gives the upper contact point $(x_1,f(x_1))$ of the metastable wall
with the top border $f(x)$. The depinning field is
 $H_c$, above which there are no metastable states.}
\label{fig:coseno}
\end{center}
\end{figure}

As an illustration consider the geometry displayed in
Fig.~\ref{fig:coseno}, $f(x)=2 - \cos(x)$. As a function of the
field $H$ stable arcs have a radius $r=\sigma/2H$ and the upper
contact point must satisfy
\begin{equation}
r = \frac{2-\cos(x)}{\sin(x)}\sqrt{1+\sin(x)^2}
\end{equation}
The solutions of this equation for different fields are shown
graphically in Fig.~\ref{fig:coseno} (upper plot), and the
corresponding arcs are shown in Fig.~\ref{fig:coseno} (lower
plot). The critical state (also shown in Fig.~\ref{fig:coseno}
(lower plot)) has $x_c^{\rm up} \approx 0.85$ and corresponds to
$r_c \approx 2.23$ and $x_2^{\rm low} \approx 1.30$. The initially
flat interface for $H=0$ will shift forward quasistatically upon
increasing the field, following the $x_1(r)$ curve. Above
$H_c=\sigma/(2r_c)$ the interface will move at a finite speed.

\begin{figure}
\begin{center}
\includegraphics[angle=0, width=7cm]{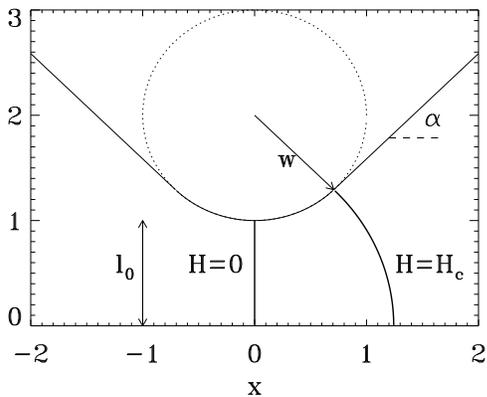}
\caption{A tip rounded at the scale $W$.} \label{fig:tip2}
\end{center}
\end{figure}

Let us now analyze the depinning from a rounded tip of curvature
radius $W$, as shown in Fig.~\ref{fig:tip2}. We assume, for
concreteness, the form
\begin{eqnarray}
f(x)&=&(l_0/2+W)-\sqrt{W^2-x^2}\;\;,|x|< x_0\\
f(x)&=& f(x_0) + \tan \alpha |x - x_0|\;\;, |x|> x_0
\end{eqnarray}
where $x_0 = W \sin(\alpha)$ and $f(x_0)=(l_0/2+W)-W \cos \alpha$. The
first equation describes a rounded circular point, and the second a
line with the asymptotic slope angle $\alpha$. In this case $x_1$
increases monotonically from zero and no more solutions of
Eq.~(\ref{solucion}) exist for $x > x_0$. We thus have $ x_c = W\sin
\alpha $ and  $r_c = f(x_0)/\sin \alpha$. The critical field is
therefore
\begin{equation}
 H_c = \frac{\sigma \sin \alpha}{l_0 + 2W(1- \cos \alpha)}.
\label{hc}
\end{equation}
For a sharp $W\to0$  tip we have
\begin{equation}
H_c = \frac{\sigma \sin \alpha}{l_0}. \label{eq:hctilt}
\end{equation}
implying a very strong pinning in the limit of strong constriction
$l_0 \to 0$. Interestingly, in this limit the depinning field would
be ultimately controlled by the rounding $W$ in more realistic
rounded tips.

\section{Building a 2D ratchet}
\label{sec:ratchet}

\begin{figure}
\begin{center}
\includegraphics[angle=0, width=7.5cm]{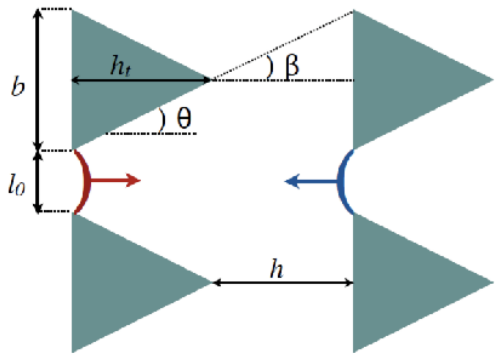}
\caption{(color online) A two dimensional ratchet  geometry made by triangles.}
\label{fig:triangles0}
\end{center}
\end{figure}

From this general theory of interface pinning, we will show how to
build a 2D ratchet for extended domain walls with both direct and
inverted rectification effects, as a function of the applied
field, by choosing the appropriate geometry for an array of
asymmetric holes. We take as a starting point the geometry
depicted in Fig.~\ref{fig:triangles0}, where triangular defects
are distributed in a rectangular array, which is similar to the
hole arrangement analyzed in Ref.~\cite{prl}. This array of
triangles presents a symmetry of reflection along the X axis but
broken reflection symmetry along the Y axis, which is the basic
condition for the observation of ratchet effects: the equivalence
between forward and backward domain wall propagation is broken in
the array allowing for domain wall rectification effects.

\begin{figure}
\begin{center}
\includegraphics[angle=0, width=8.5cm]{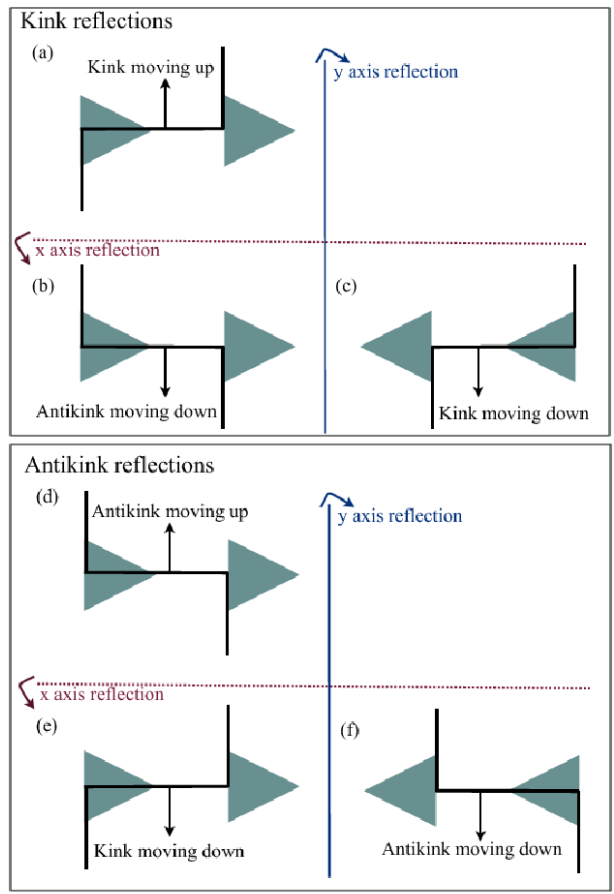}
\caption{(color online) Broken symmetry in a 2D kinked wall. (a)
Kink moving up. (b) After an x-axis reflection, an antikink moving
down is obtained. (c) After a second y-axis reflection, a kink
moving down on a different array (inverted triangle array) is
obtained. From (d) to (f) same reflections are shown for an
antikink. } \label{fig:broken2d}
\end{center}
\end{figure}

This kind of ratchet effect has been mostly studied in 1D magnetic
nanowires \cite{notches1,notches2,angelfish,allwood}. However, in a 2D array such as shown in
Fig.~\ref{fig:triangles0} the 1D character of the elastic domain walls
opens the possibility of extra propagation modes. In particular,
when a wall is pinned between two lines of defects, it develops
kinks and antikinks, as shown in Fig.~\ref{fig:broken2d}.
Depending on their shape and the sign of the field, kinks and
antikinks can move upward or downward, turning in a net wall
motion to the left or to the right. This kink motion is also
asymmetric, reflecting the Y-axis asymmetry of the pinning
potential by the array of triangles, so that it opens the
possibility of a rectified motion of the kinked wall. In
particular, let us analyze in detail how the symmetry properties
of the array influence kink propagation: for example, as shown in
Fig.~\ref{fig:broken2d}, a kink moving upward
(Fig.~\ref{fig:broken2d}(a)) is equivalent to an antikink moving
downward (Fig.~\ref{fig:broken2d}(b)) upon reflection along the
X-axis, which is an allowed symmetry operation of the array of
triangles. In fact, both movements (kink upward and antikink
downward) result in a net backward motion of the extended domain
wall. The critical field for this propagation process will be
labelled $H_{\rm U}$ from now on. However, a kink moving downward
(Fig.~\ref{fig:broken2d}(c)) is the result of a reflection upon
the Y axis of the antikink moving downward. This is a broken
symmetry in the array (note the inverted triangles), implying that
both situations are not equivalent. Actually, both the downward
motion of a kink (Fig.~\ref{fig:broken2d}(e)) and the upward
motion of an antikink (Fig.~\ref{fig:broken2d}(d)), which are
equivalent upon reflection along the X-axis, result in a net
forward motion of the extended wall. The critical field for this
propagation process will be labelled $H_{\rm D}$ in the following. In
short, $H_{\rm U}$ and $H_{\rm D}$ could not be the same due to the broken Y
axis symmetry in the array.

Thus, to understand domain wall propagation in the rectangular
array of triangles two facts must be considered: first, the broken
Y-axis symmetry breaks the equivalence between forward and
backward domain wall propagation; second, the extended nature of
domain walls in the 2D array of holes allows for extra propagation
modes not possible in 1D geometries, such as those corresponding to nanowires:
flat wall propagation and kinked
wall propagation. If a wall is pushed to the right by the applied
field two different propagation modes can be activated: either
forward flat wall propagation (at a critical field $H_{\rm F}$) or
kinked wall propagation (by kinks moving downward and/or antikinks
moving upward at $H_{\rm D}$). On the contrary, if a wall is pushed to
the left by the applied field the possible propagation modes will
be either backward flat wall propagation (at a critical field
$H_{\rm B}$) or kinked wall propagation (by kinks moving upward and/or
antikinks moving downward at $H_{\rm U}$). The global behavior of walls
upon propagation across the array of asymmetric defects will
depend on the relationships between the four relevant critical
fields $H_{\rm F}$, $H_{\rm B}$, $H_{\rm U}$ and $H_{\rm D}$. Actually, for some
geometries, it can happen that the rectification of kink motion is
opposite to the rectification of a vertical non-kinked wall
described previously. This crossed ratchet effect offers promising
technological applications since it allows a non trivial control
of the two dimensional wall.

First, we will analyze the propagation of a flat wall crossing a
line of triangular defects (forward-backward ratchet), which is
equivalent to the propagation of domain walls in nanowires with
asymmetric geometry \cite{notches1,notches2,angelfish,allwood}.
Then, we will study the
upward-downward propagation of a kink in a wall pinned in between
two adjacent defect lines (upward-downward ratchet). Finally, we
will discuss the geometrical parameters needed to design 2D arrays
of asymmetric holes with crossed ratchet behavior (opposite sign
for forward-backward and upward-downward ratchets).

\subsection{Flat walls: Forward-Backward propagation}
\label{sec:finitewidth}

An infinitely narrow domain wall moving from the left to the right
(forward) across a vertical line of triangles  (see Fig.~\ref{fig:triangles0})
will be pinned at the
base of the triangles, where the distance between the ends of the wall is minimum.
The depinning field can be derived from
the one obtained in subsection \ref{sec:depinning} for a geometry defined tip and a flat boundary, as in Fig.~\ref{fig:tip2}. Now we have two symmetric tips but this situation is equivalent to the previous one if we add the mirror image of the tip, the wall adopting the same shape as the one depicted in Fig.~\ref{fig:tip2} (plus its mirror image). Therefore, the depinning field is identical to the one given by
Eq.~(\ref{eq:hctilt}):
\begin{equation}
H_{\rm F}=\frac{\sigma \sin\theta}{l_0} \label{eq:hrnarrow}
\end{equation}
where $\theta$ is the angle between the sides of the triangles and
the horizontal, and  $l_0$ is the vertical distance
between triangles (see Fig.~\ref{fig:triangles0}). Notice also that the critical field (\ref{eq:hrnarrow}) is, according to Eq.~(\ref{anglecondition}), the one for which the wall accommodates to the boundary of both triangles.

A wall moving to the left (backward), like the one depicted in
Fig.~\ref{fig:triangles0}, will be also pinned between the same
vertices of the triangles, but now it has to grow along the vertical
bases, i.e., the angle of the boundaries in Eq.~(\ref{eq:hctilt}) is
$90^{\rm o}$. Therefore:
\begin{equation}
H_{\rm B}=\frac{\sigma}{l_0} \label{eq:hlnarrow}
\end{equation}

We find $H_{\rm F}/H_{\rm B} = \sin\theta \leq 1$, i.e., it is
easier for the wall to move forward than backward, as
expected. The triangles can therefore rectify the motion of the
wall. Applying an alternating field of peak intensity $H$, with
$H_{\rm B}<H<H_{\rm F}$, the wall will have a net forward motion, so that a direct ratchet effect is obtained.

\begin{figure}
\begin{center}
\includegraphics[angle=0, width=8.5cm]{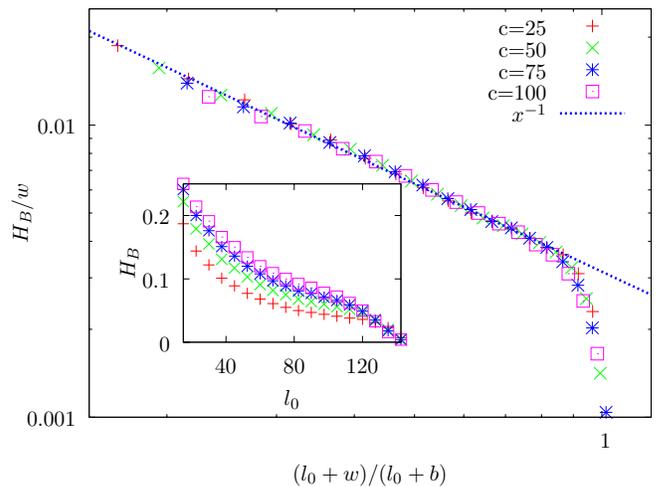}
\caption{(color online) Scaled $H_{\rm B}/w$ vs. $(l_0+w)/(l_0+b)$
for different values of $c$, $l_0$ and $b$. Solid line indicates
the $H_{\rm B}/w \sim (l_0+w)^{-1}$ dependence. Inset shows
backward depinning field for a $\phi^4$ domain wall, $H_{\rm B}$,
as a function of the vertical gap between triangles, $l_0$, for
different values of the elastic constant $c$ or domain wall width
$w$.} \label{fig:fig1_phi4}
\end{center}
\end{figure}

\begin{figure}
\begin{center}
\includegraphics[angle=0, width=8.5cm]{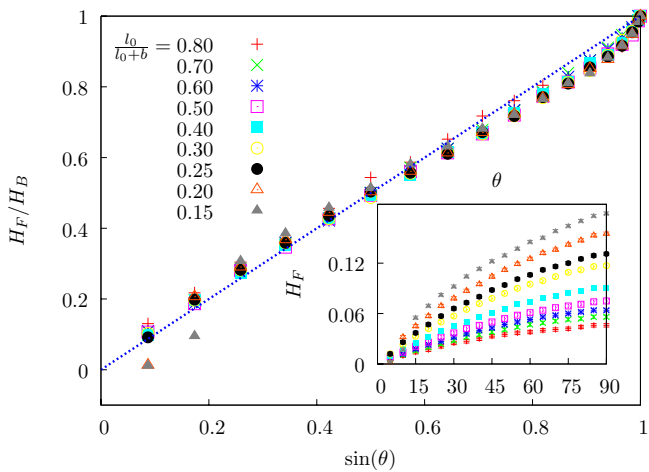}
\caption{(color online) Forward/backward asymmetry for the flat
$\phi^4$ wall with $c=50$, $H_{\rm F}/H_{\rm B}$ vs. $\sin
\theta$, calculated for different $l_0/(l_0+b)$. Inset shows the
forward depinning field, $H_{\rm F}$ for a $\phi^4$ domain wall,
as a function of the isosceles angle, $\theta$, for different
values of the vertical gap between triangles, $l_0$.}
\label{fig:fig2_phi4}
\end{center}
\end{figure}

\textbf{Finite width effects.} We compare our simple previous
geometrical estimates for $H_{\rm F}$ and $H_{\rm B}$ with
simulations for a more realistic $\phi^4$ domain wall with a
finite width, as it was done in Ref. [\onlinecite{prl}]. In the
inset of Fig.~\ref{fig:fig1_phi4} we plot the backward depinning
field, $H_{\rm B}$, of a $\phi^4$ domain wall as a function of
$l_0$ for different elastic constants $c$ and fixed $\epsilon_0
=1$. This is equivalent to change the  domain wall width  $w =
\sqrt{2c/\epsilon_0}$ and the domain wall energy $\sigma =
\sqrt{8c\epsilon_0}/3$ that scales as $\sigma \sim w$. $H_{\rm B}$
is found to increase as a function of $c$, mainly due to the
increase in domain wall energy $\sigma$ and to decrease as a
function of $l_0$. All the data of $H_{\rm B}$ obtained from the
different simulations can be scaled to a single curve, if we plot
$H_{\rm B}/w \sim H_{\rm B}/\sigma$ as a function of
$(l_0+w)/(l_0+b)$ (see main panel of Fig.~\ref{fig:fig1_phi4}).
For small values of $l_0$, $H_{\rm B}/w \sim (l_0+w)^{-1}$ as
predicted by Eq.~(\ref{eq:hlnarrow}) except for a correction to
$l_0$ which is of the order of the domain wall width $w$. This
correction can be qualitatively understood by noting that the
$\phi^4$ wall just below $H_{\rm B}$ extends from the tip of the
triangles up to a distance of order $w$ into the base of each of
them, so that the center of the wall describes an arc covering a
vertical distance $l_0+w$. It is interesting to mention that this
correction is the same as predicted by Eq.~(\ref{hc}) for a
rounded tip of curvature radius equal to domain wall width ($W=w$)
and $\alpha = 90^{\rm o}$ (as corresponds for backward depinning).
That is, finite domain wall width and tip rounding of defect
geometry have equivalent effects on depinning fields, softening
the magnetic behavior in comparison to analytical calculation for
sharp tips and narrow domain walls. At large values of $l_0$,
$H_{\rm B}/w$ deviates from the behavior $H_{\rm B}/w \sim
(l_0+w)^{-1}$ decreasing steeply precisely when
$(l_0+w)/(l_0+b)\sim 1$. The reason is that for the simulations
with varying $l_0$ we fix the periodicity of the lattice $l_0+b$.
Therefore, for large values of $l_0$ the base of triangles becomes
small, and eventually of order $w$, strongly reducing the
geometric pinning mechanism when $w\sim b$.

In the inset of Fig.~\ref{fig:fig2_phi4} we show $H_{\rm F}$ vs.
$\theta$ for the $\phi^4$ wall, for
different values of $l_0$ and constant $c = 50$. The main panel of
Fig.~\ref{fig:fig2_phi4}, shows the forward/backward asymmetry for
the flat wall calculated as $H_{\rm F}/H_{\rm B}$ in comparison
with the analytical prediction of Eq.~(\ref{eq:hrnarrow}) for a
narrow domain wall $H_{\rm F}/H_{\rm B}= \sin \theta$. The
simulated values follow nicely the $\sin \theta$ line except for
small deviations at small and large angles $\theta\sim90^{\rm o}$. These
can be in part attributed to the discreteness of the lattice,
which does not allow to produce smooth slopes at the scale $w$
when the angle is too close to $\theta=0$ and $\theta=90^{\rm o}$.

In short, our simulations with the $\phi^4$ model are consistent
with the geometric estimates for a narrow wall and show how to
correct the depinning fields for single arcs with a finite width,
which can be relevant for experimental situations \cite{JPD}. The depinning
of single arcs are, on the other hand, the main building blocks
for calculating all the anisotropic depinning fields and, in
particular, the crossed-ratchet effect. Thus, the softening of the
critical fields observed due to finite width corrections and/or
the effect of rounded tips could also be applied in a similar way
to the geometric estimates of the propagation of kinked walls.

\begin{figure}
\begin{center}
\includegraphics[angle=0, width=7cm]{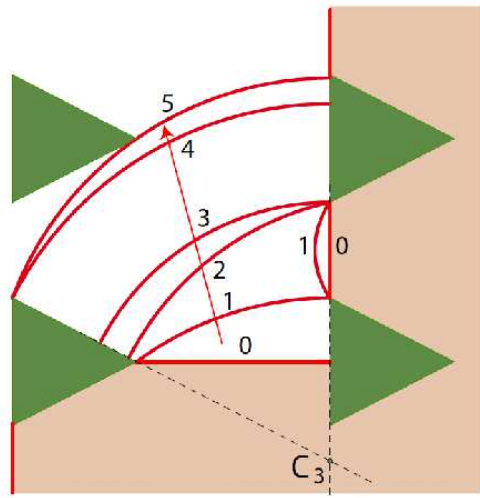}
\caption{(color online) Upward motion of a kink in an elastic
wall (red online).} \label{fig:up}
\end{center}
\end{figure}

\subsection{Kinked wall: Upward-Downward propagation} \label{sec:crossedratchets}

\textbf{Upward propagation.} The depinning field for kinks can
also be calculated using the basic geometry analyzed in
Sec.~\ref{sec:basic}. Fig.~\ref{fig:up} shows the evolution of
a wall forming a kink as it is pushed upwards  when the field
increases from $H=0$. The initial disposition of the wall is
labelled as $0$. The critical field $H_{\rm U}=\sigma/(2r_{\rm
min})$ for a complete depinning of the kink is given by the arc
with minimal radius $r_{\rm min}$. For a given geometry, one has
to carefully trace the trajectory of the wall, as depicted in
Fig.~\ref{fig:up}, and compute the minimal radius in each step.

To step from 0 to 1, i.e. to depin the transverse horizontal
segment of the kink, is similar to the situation depicted in
Fig.~\ref{fig:tip2} with $\alpha=90^{\rm o}-\theta$. Therefore,
the corresponding critical field is $\sigma\cos\theta/(2h)$.
However, the base $b$ of the triangle can be too short for the
domain wall to reach the symmetrical position described in
Fig.~\ref{fig:tip2}. The wall is maximally tilted at an angle
$\beta$ (see Fig.~\ref{fig:triangles0}). In this situation, the
corresponding angles $\theta_{1}=\theta_{2}$ in Fig.~\ref{fig:arc}
are equal to $\beta$ (the angle formed with the base of the
rightmost triangle in Fig.~\ref{fig:up}) and the distance
between the two contact points is $l=h/\cos\beta$. The wall
reaches this orientation  if $H>\sigma
\sin\beta/l=\sigma\sin(2\beta)/(2h)$. Consequently, the critical
field to move from 0 to 1 is the minimum of these two fields,
namely,
\begin{eqnarray} H_{\rm 1}^{\rm U} &=& \frac{\sigma}{2h}\min \left\{
\cos(\theta),\sin(2\beta) \right\} \nonumber \\
&=&\frac{\sigma}{2h}\min \left\{
\cos(\theta),\frac{bh}{h^2+(b/2)^2}\right\}
\end{eqnarray}

The next critical arc is 3, a wall orthogonal to the contact sides
of the triangles across a diagonal of the rectangular cell of
triangles. The center of this arc is the point $C_{3}$, the
intersection between the two prolongations of the triangles sides.
The radius of arc 3 is $l_{0}+b/2+h/\tan\theta$, and the
corresponding critical field:
\begin{equation}
H^{\rm U}_{\rm 3}=  \frac{\sigma}{2h}  \left( \frac{1}{l_0/h
+\tan\beta+1/\tan\theta} \right) 
\end{equation}

 Finally, one should also consider the diagonal arc 5 to complete the upward motion of the kink.
 However, the radius of this arc is bigger than $h_{t}+h=b/(2\tan\theta)+h$, resulting
 in lower critical fields for the geometries considered in this paper. The final result for upward motion is:
 \begin{equation}
H_{\rm U}=\max\{H^{\rm U}_{1},H^{\rm U}_{3}\}
\end{equation}

\begin{figure}
\begin{center}
\includegraphics[angle=0, width=8.2cm]{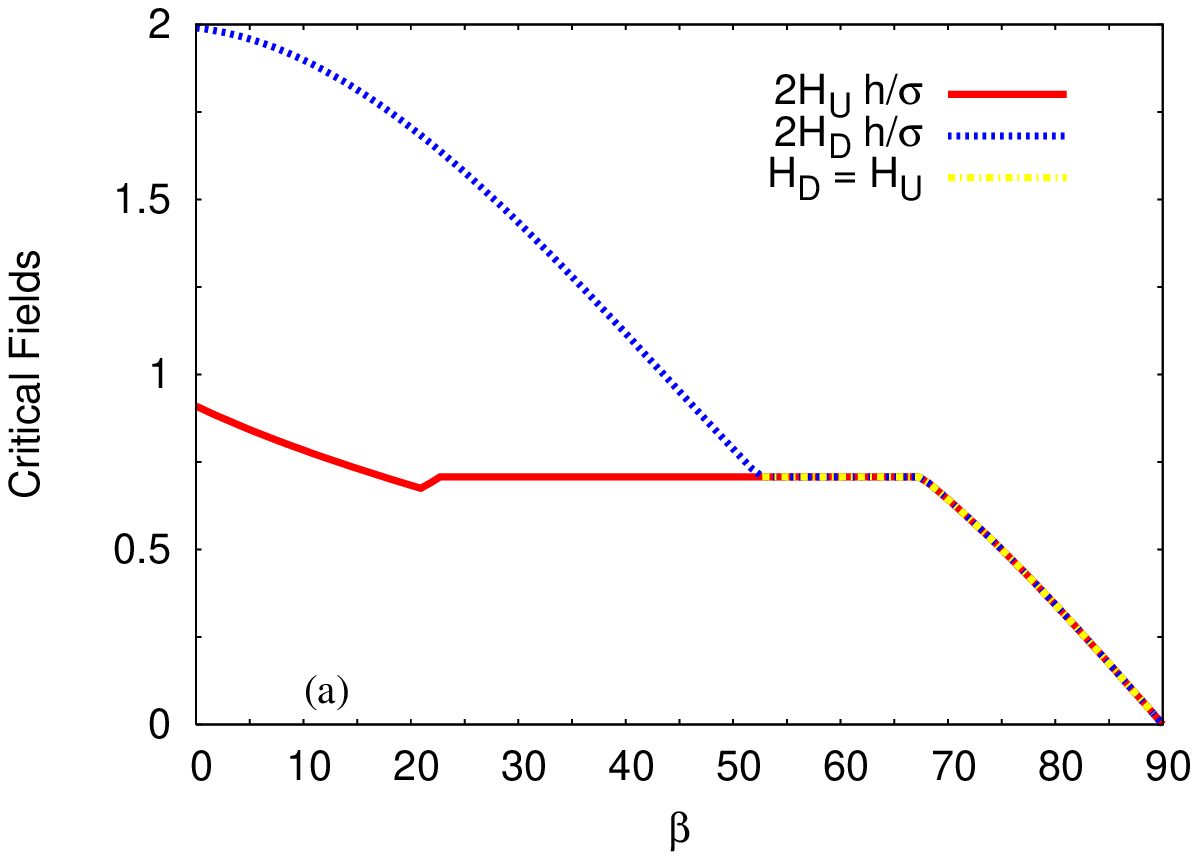}
\includegraphics[angle=0, width=8.2cm]{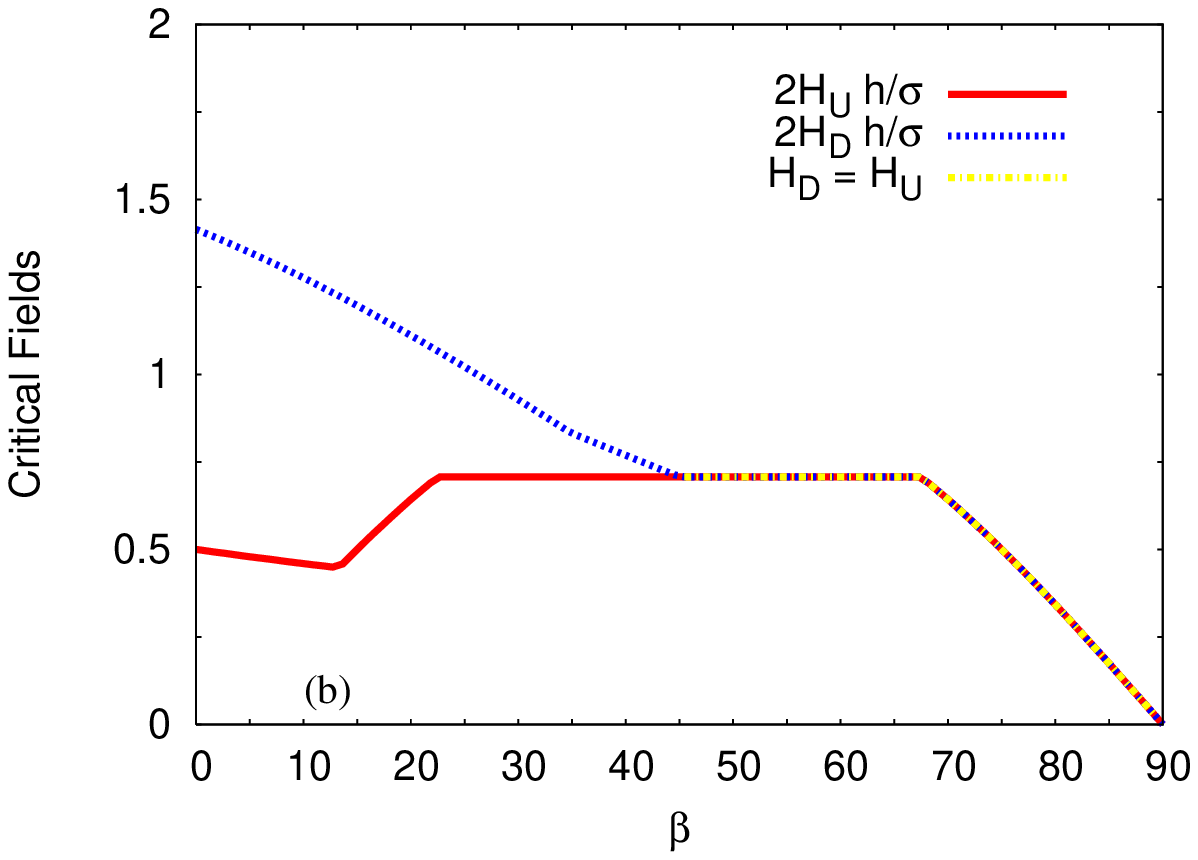}
\includegraphics[angle=0, width=8.2cm]{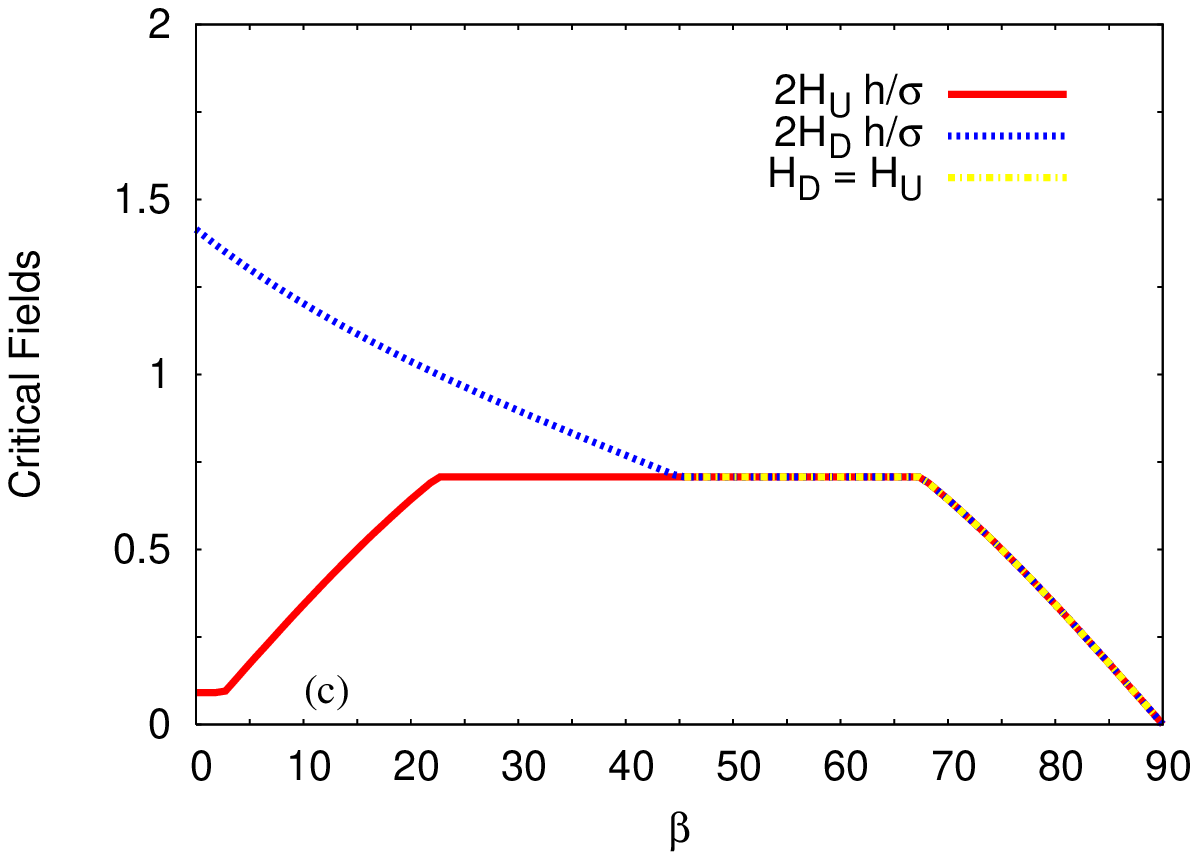}
\caption{Critical fields, $H_{\rm D}$ and $H_{\rm U}$ normalized
by $\sigma/2h$ vs. $\beta$ calculated for $\theta = 45^{\rm o}$
for (a) $h/l_0 = 10$ (b) $h/l_0 = 1$ and (c) $h/l_0 = 0.1$. Note
that for large $\beta$, $H_{\rm D}$ becomes equal to $H_{\rm U}$
and, therefore, a single line appears in the figures.}
\label{fig:crifields1}
\end{center}
\end{figure}

From the above equations, three geometrical parameters of the
rectangular array of triangles are found to control de interplay
between the different depinning processes of the kinked wall and,
thus, the relevant critical fields: the angle $\theta$ that
defines triangle shape, the angle $\beta$ that characterizes the
shape of the horizontal intertriangle region ($\beta$ is given by
$\tan \beta=b/2h$), and the ratio $h/l_0$ between horizontal and
vertical triangle distance. This last parameter, $h/l_0$, is only
important in the depinning of the diagonal arc 3. Figure
\ref{fig:crifields1} shows the calculated $H_{\rm U}$, normalized
by the scale factor $\sigma/2h$, as a function of $\beta$ for
$\theta = 45^{\rm o}$ and different values of the ratio $h/l_0 =
10, 1, 0.1$. For large $\beta$, $H_{\rm U}$ is given by $H^{\rm
U}_{1}$, so that it is the same in the three panels of Fig.
\ref{fig:crifields1}. Below $\beta \simeq 22^o$, there is a
crossover to $H_{\rm U} = H^{\rm U}_{3}$ indicated by the upturn
in $H_{\rm U}(\beta)$ as $\beta$ decreases. It occurs at different
angular positions depending on $h/l_0$: $\beta_c = 2.6^o, 13^o$
and $21^o$ for $h/l_0 = 0.1, 1, 10$, respectively. That is, for
small $\beta$ and large $h/l_0$ (very anisotropic rectangular
array), critical upward depinning occurs at the diagonal arc 3 in
Fig.~\ref{fig:up}, whereas in the rest of the parameter space
the relevant process corresponds to depinning of the transverse
domain wall segment between adjacent triangles in the same row
(arc 1 in Fig.~\ref{fig:up}).

\begin{figure}
\begin{center}
\includegraphics[angle=0, width=7cm]{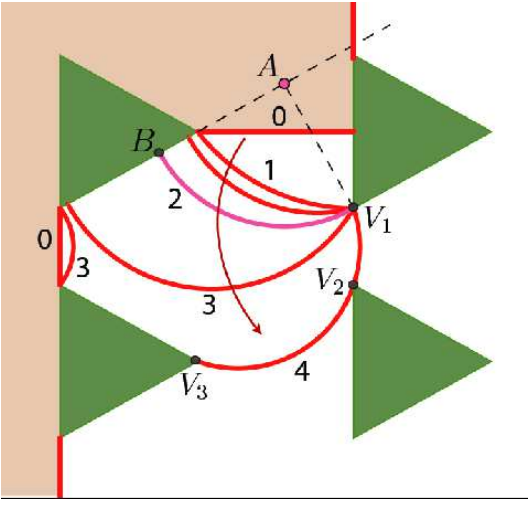}
\caption{(color online) Downward  motion of a kink in an elastic
wall (red online).} \label{fig:down}
\end{center}
\end{figure}

\textbf{Downward propagation.} The first step $0\to 1$ of the
downward motion (Fig.~\ref{fig:down}) is identical to the same
step in the upward motion, hence $H^{\rm D}_{1}=H^{\rm U}_{1}$.
From 1 to 3, the arc 2 has the minimal radius, which equals the
distance between its center $A$ and the vertex $V_{1}$,
$(h+h_{t})\sin\theta$. However, point $B$ can lie below or above
the side of the triangle. In the first case, which occurs if
$\theta>45^{\rm o}$, the critical field is given by the radius of
arc 3, $(h +h_{t})/(2\cos\theta)$. The second case occurs when
$\theta<45^{\rm o}-\beta$ and then the minimum radius,
$h/(2\cos(\beta+\theta)\cos\beta)$ occurs when point B is at the
vertex of the triangle. The field to reach arc 3 can be written
as:
\begin{equation}
H^{\rm D}_{3}=\frac{\sigma}{2h} \left\{\begin{array}{ll}
\displaystyle  2\cos(\beta+\theta)\cos\beta & \mbox{if
$\theta<45^{\rm o}-\beta $}\\  & \\\displaystyle
\frac{1}{[\tan\beta/\tan\theta+1]\sin\theta}  &
\mbox{if $45^{\rm o}-\beta<\theta<45^{\rm o}$}\\ & \\
\displaystyle \frac{2\cos\theta}{\tan\beta/\tan\theta+1} &
\mbox{if $\theta>45^{\rm o}$}
\end{array}\right.
\end{equation}

Finally, the radius of the diagonal arc 4 is
\begin{equation}
r_{4}=\frac{1}{2}\sqrt{\left( h
+\frac{2l_{0}b+b^{2}}{4h}\right)^{2}+l_{0}^{2}}
\end{equation}
yielding
\begin{equation}
H^{\rm D}_{4}= \frac{\sigma}{2r_{4}} \nonumber \\
\end{equation}

\begin{equation}
= \frac{\sigma}{2h}
\left( \frac{2}{\sqrt{(l_0/h)^2+(1+l_0/h \tan\beta + \tan^2\beta)^2}} \right) \\
\end{equation}
and
\begin{equation}
H^{\rm D}=\max\{H^{\rm U}_{1},H^{\rm D}_{3},H^{\rm D}_{4}\}
\label{eq:HD}
\end{equation}

Figure \ref{fig:crifields1} shows the calculated $H_{\rm D}$,
normalized by the scale factor $\sigma/2h$, as a function of
$\beta$ for $\theta = 45^{\rm o}$ and different values of the
ratio $h/l_0 = 10, 1, 0.1$. For large $\beta$, $H_{\rm D}$ is
given by $H^{\rm D}_{1}$, i.e. depinning of the transverse
horizontal wall segment, but as $\beta$ decreases $H^{\rm D}_{3}$
and $H^{\rm D}_{4}$ become more relevant. In particular, for large
$h/l_0$, $H^{\rm D}_{4}$ dominates the behavior in a wide $\beta$
range. The result is that, in the low $\beta$ range, $H_{\rm D}$
is much larger than $H_{\rm U}$, but above a certain threshold
$\beta_0$ both fields become equal ($H_{\rm D} = H_{\rm U}$) (for
example, for $h/l_0 = 1$, $\beta_0 = 45^o$). Thus, for
$\beta<\beta_0$, upward kink propagation is easier than downward
propagation so that when an alternating field of peak intensity
$H$, with $H_{\rm U}<H<H_{\rm D}$ is applied to the kinked wall,
it will have a net backward motion, i.e. opposite to the behavior
observed in the previous subsection on flat wall propagation
walls. On the other hand, for $\beta>\beta_0$, $H_{\rm D} = H_{\rm
U}$, so that kink propagation is not rectified by the array of
triangles.

\begin{figure}
\begin{center}
\includegraphics[angle=0, width=8cm]{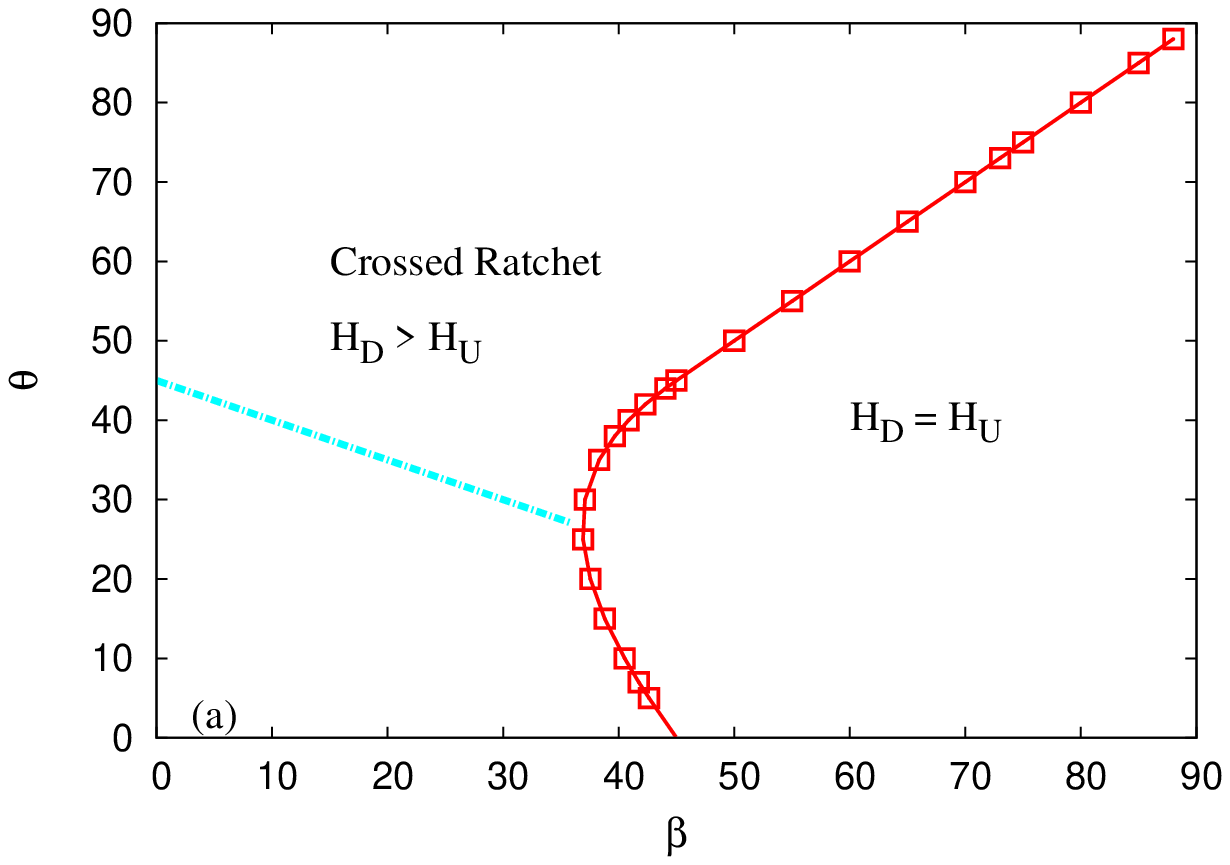}
\includegraphics[angle=0, width=8cm]{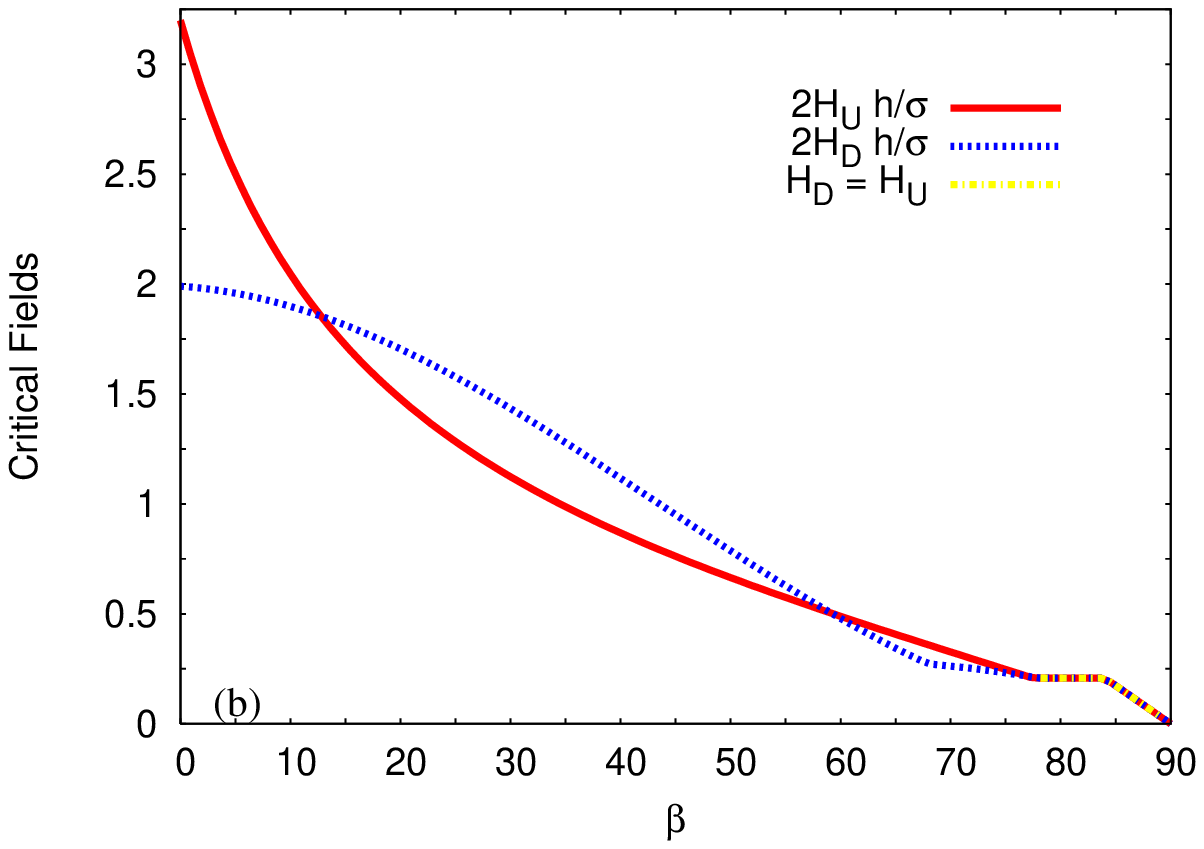}
\includegraphics[angle=0, width=8cm]{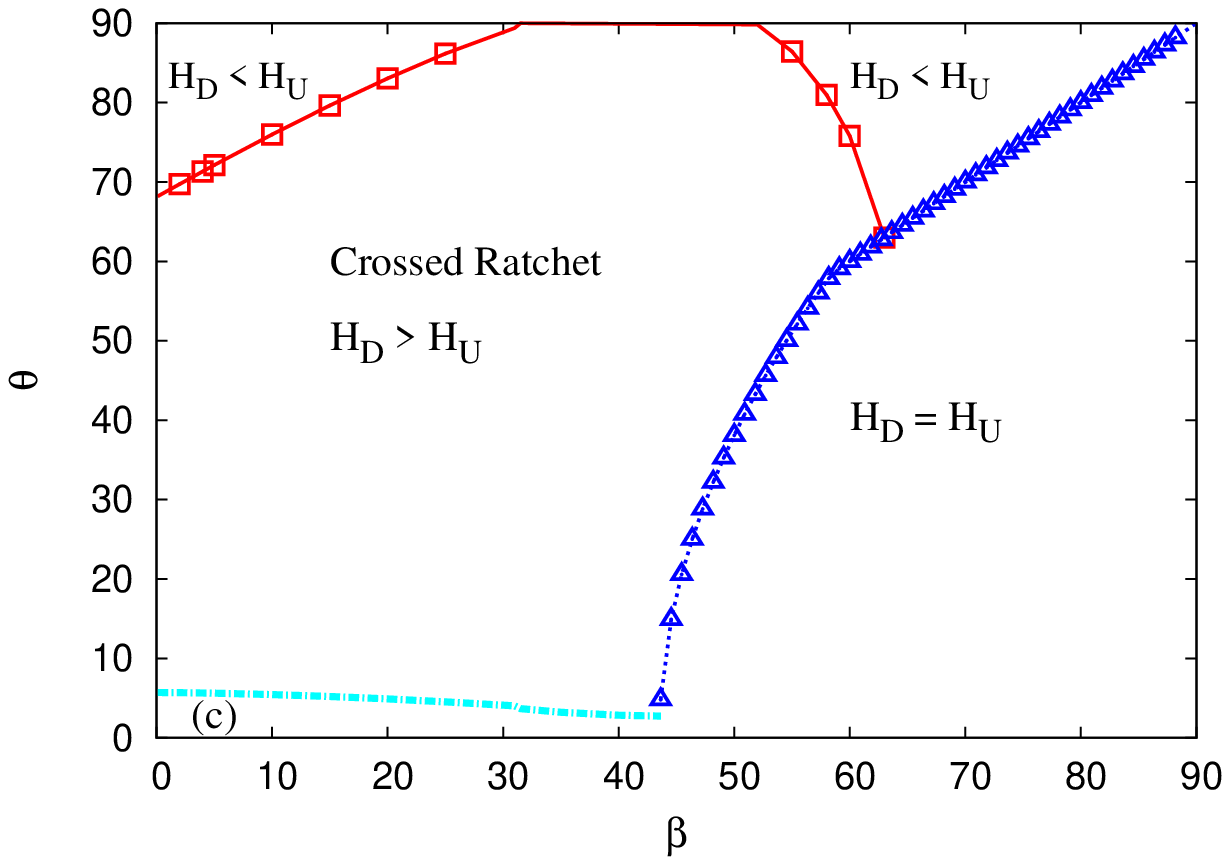}
\caption{(color online) (a) Phase diagram in ($\beta, \theta$)
plane for of the different regimes for kinked wall propagation for
$h/l_0 = 1$; (b) Critical fields, $H_{\rm D}$ and $H_{\rm U}$
normalized by $\sigma/2h$ vs. $\beta$ calculated for $\theta =
78^{\rm o}$ for $h/l_0 = 10$; (c) Same as in (a) for $h/l_0 = 10$.
Dotted lines in (a) and (c) indicate the condition $H_{\rm D} =
H_{\rm F}$ below which flat wall propagation modes compete with
kinked wall propagation.
 } \label{fig:crifields2}
\end{center}
\end{figure}

In short, for $\theta = 45^0$, as is the case in
Fig.\ref{fig:crifields1}, whenever kinked wall propagation is
asymmetric, it results in an inverted ratchet effect. This is
actually the case for most of the parameter space ($\beta, \theta,
h/l_0$). For example, Fig.~\ref{fig:crifields2}(a) shows the
phase diagram in the ($\beta, \theta$) plane for $h/l_0 = 1$, in
which only these two regimes for domain wall propagation are
found: inverted ratchet ($H_{\rm D} > H_{\rm U}$) in the low
$\beta$ region and symmetric kink propagation in the right bottom
corner of the diagram. This is a direct consequence of the maximum
condition imposed in eq.~(\ref{eq:HD}), as long as $H^{\rm U} =
H^{\rm U}_{1}$. However, at large $h/l_0$, the role of arc 3 in
Fig.~\ref{fig:up} in critical upward depinning becomes more
important and $H^{\rm U}$ is given by $H^{\rm U}_{3}$ in a wider
($\beta, \theta$) region. In this case, $H_U$ can take any value
in comparison with $H_D$, so that a direct ratchet effect for
kinked wall motion becomes possible. An example of this situation
can be seen in Fig.~\ref{fig:crifields2}(b) for $\theta =78^o$ and
$h/l_0 = 10$. Thus, the phase diagram for this very anisotropic
array of triangles with $h/l_0 = 10$ (Fig.~\ref{fig:crifields2}
(c)) becomes more complex: inverse ratchet effect ($H_{\rm D} >
H_{\rm U}$) is found in a large ($\beta, \theta$) region in the
left part of the diagram, kink motion is symmetric in the right
part of the diagram ($H_{\rm D} = H_{\rm U}$) and, finally, direct
ratchet (i.e. $H_{\rm D} < H_{\rm U}$) is found in two small
regions close to the upper part of the diagram, above $\theta =
65^o$. This direct ratchet regions shrink as $h/l_0$ decreases and
disappear for $h/l_0 < 2$ due to the softening of the depinning
processes of the diagonal arcs in comparison with depinning of the
horizontal transverse segments.

\subsection{Crossed ratchets}

From the previous analysis, we have found a fundamental difference
between flat and kinked  wall propagation modes: $H_{\rm F}/H_{\rm
B}$ is always smaller than unity, implying that flat wall
propagation under an ac field will result in direct ratchet
effects; on the contrary, $H_{\rm U}/H_{\rm D}$ can take any value
so that kinked wall motion can result either in direct and inverse
rectification effects. Thus, the first condition to design an
asymmetric array of defects that displays crossed ratchet behavior
is to choose a point in the phase diagram of 
Fig.~\ref{fig:crifields2} in which $H_{\rm U}/H_{\rm D}<1$. Then,
in order to observe clear crossed ratchet effects that can be
useful for device applications, the interplay between the four
relevant critical fields $H_{\rm F}$, $H_{\rm B}$, $H_{\rm D}$ and
$H_{\rm U}$ must be taken into account.

\begin{figure}
\begin{center}
\includegraphics[angle=0, width=8cm]{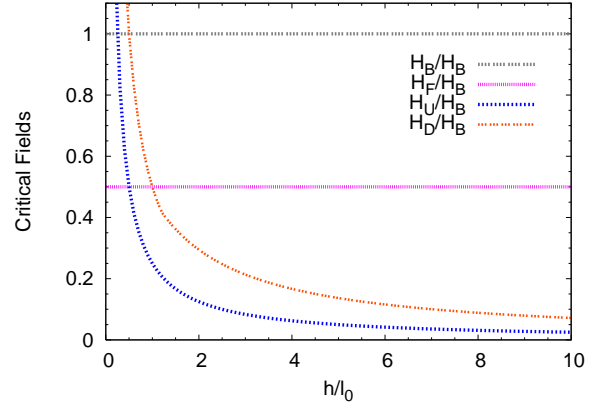}
\caption{(color online) Critical fields, for $\theta=30^{\rm o}$,
and $\beta=30^{\rm o}$ as a function of $h/l_0$. }
\label{fig:crifields}
\end{center}
\end{figure}

In Fig.~\ref{fig:crifields} we plot the four critical fields as a
function of $h/l_0$ between triangles, normalized by $H_{\rm B}$
for $\beta = 30^o$ and $\theta =30^{\rm o}$, which is similar to
the geometry used in the experiments of Ref.~\onlinecite{prl},
where crossed ratchet effects were observed both experimentally
and theoretically. The crossed ratchet effect is apparent from the
figure: $H_{\rm F}$ is smaller than $H_{\rm B}$, but $H_{\rm U}$
(the upward motion of the kink drives the wall backwards) is
larger than $H_{\rm D}$. Domain wall propagation in the array is
determined by the relationships between the four critical fields:
for each particular array geometry (i.e. each $h/l_0$ value) when
a domain wall is pushed in the forward direction it will depin as
soon as the applied field reaches the lowest of $H_{\rm F}$ and
$H_{\rm D}$; but, when a domain wall is pushed in the backward
direction it will depin as soon as the applied field reaches the
lowest of $H_{\rm B}$ and $H_{\rm U}$. For large $h/l_0$ (i.e.
very rectangular array cell), both $H_{\rm D}$ and $H_{\rm U}$
take much lower values than $H_{\rm B}$ and $H_{\rm F}$ implying
that two well separated field ranges can be defined: low field
domain wall propagation dominated by kink motion (i.e. easier
backward wall motion) and high field domain wall propagation
dominated by flat wall motion (i.e. easier forward wall motion).
As $h/l_0$ is reduced below $\approx 1$, $H_{\rm D}$ becomes
larger than $H_{\rm F}$, the interplay between flat and kinked
wall propagation modes becomes more complex, and the different
rectification effects cannot be clearly separated. Finally, for
very close triangle lines ($h/l_0$ below $0.25$), $H_{\rm U}$
becomes larger than $H_{\rm B}$ and domain wall motion in the
array is dominated by flat wall propagation modes.

The condition $H_{\rm D} = H_{\rm F}$ is plotted as a dotted line
in Figs.~\ref{fig:crifields2} (a) and (c), so that the region
for well separated kinked and flat propagation modes, i.e. clear
crossed ratchet observation, lies above this line in the ($\beta,
\theta$) plane. It can be seen that as $h/l_0$ increases the
available parameter region for crossed ratchet becomes wider due
to the different scaling of the critical fields: $H_{\rm F}$ and
$H_{\rm B}$ scale as $1/l_0$, whereas $H_{\rm D}$ and $H_{\rm U}$
scale as $1/h$. Thus, the design of arrays in the large $h$ range
($h \gg l_0$) appears as an important condition for a clear
observation of crossed ratchet effects that can be of use in
device applications.

\section{Conclusions}
\label{sec:conclusions}

In summary, the propagation of an elastic domain wall in a two
dimensional medium has been analyzed in an arbitrary geometry
defined by holes and sample boundaries. The local depinning fields
for an anchored wall have been calculated as a function of
boundary shape in terms of the minimal arc radius that satisfies
the relevant orthogonality conditions. Then, these results have
been applied to the design of 2D arrays of asymmetric holes with
broken Y-axis reflection symmetry that can display crossed ratchet
effects (i.e. direct ratchet for forward/backward flat wall
propagation and inverted ratchet for upward/downward kink
propagation).

For a rectangular array of triangles, flat wall propagation is
found to be always asymmetric resulting in a direct ratchet effect
controlled by triangle shape (angle $\theta$) and intertriangle
vertical distance ($l_0$). Corrections due to finite domain wall
width and/or rounded triangle tips, that could be relevant in
actual patterned arrays of holes, produce a global softening of
the critical fields but do not alter significantly
forward/backward asymmetry. On the other hand, upward/downward
kink propagation can display any asymmetry and depends not only on
triangle shape but also on the shape of the horizonal
intertriangle region (angle $\beta$) and on the array
vertical/horizontal anisotropy ($h/l_0$). The array geometry
needed for the observation of crossed ratchet effects has been
determined considering the different wall propagation modes
relevant in the different points of the ($\beta$, $\theta$) plane.
Anisotropic arrays with large $h/l_0$ are found to be optimum for
the observation of clear crossed ratchet effects.

\begin{acknowledgments}

Work supported by SeCyT-UNC\'{o}rdoba and  CONICET, Argentina;
Spanish MICINN (FIS2008-06249, HP2008-0032) Asturias FICYT
(IB08-106), Grant MOSAICO (Spain) and MODELICO-CM (Comunidad de Madrid, Spain). J.A. Capit\'an acknowledges funding by a
 contract from Comunidad de Madrid and Fondo Social Europeo. A.B. Kolton acknowledges ANPCYT (Grant PICT2007886, Argentina), and Universidad de
 Barcelona, Ministerio de Ciencia e Innovaci\'on (Spain) and Generalitat de
 Catalunya for partial support through I3 program.

\end{acknowledgments}

\begin{appendix}

\section{From field equations to elastic interfaces}
\label{ap:line}

In this Appendix we  construct an approximate stationary solution of
Eq.~(\ref{phi4}) around a given curve  in the plane, defined as
$(x(s),y(s))$, with $s$ a real number taking values in some
interval.

The interface of the desired solution is centered around the line
$(x(s),y(s))$, i.e., $\phi(x(s),y(s))=0$, and the field approaches
to the stable values $\pm 1$ as we move away from the line. We then
construct the solution using the {\em signed distance function},
$g(x,y)$, whose absolute value is the distance of a point $(x,y)$ to
the line $(x(s),y(s))$. Obviously, $g(x(s),y(s))=0$ for all $s$. One
less obvious property is that the gradient of the distance function
is unitary all over the plane. In other words, the distance function
obeys the {\em eikonal equation}:
\begin{equation}
\left[\partial_x g(x,y)\right]^2+\left[\partial_y
g(x,y)\right]^2=1\label{grad}
\end{equation}

Now we choose the following form for the field
$\phi(x,y)=f(g(x,y))$. Introducing this ansatz in the stationary
$\phi^4$ equation and making use of Eq.~(\ref{grad}), we get:
\begin{equation}
cf''(g)+cf'(g)\nabla^2g+\epsilon_0[f(g)-f(g))^3]=0
\label{phi01}
\end{equation}

Our first approximation consist of neglecting
$f'(g(x,y))\nabla^2g(x,y)$ in the above equation. The Laplacian of
the distance function is inversely proportional to the curvature
radius of the line $(x(s),y(s))$. Therefore, our
approximation is valid for interfaces with a curvature radius much
larger than its width. Then, Eq.~(\ref{phi01}) reduces to:
\begin{equation}
cf''(g(x,y))+\epsilon_0[f(g(x,y))-f(g(x,y))^3]=0
\end{equation}
and the general solution reads $f(z)=\tanh[(z-z_0)/w]$ with
\begin{equation}
w=\sqrt{\frac{2c}{\epsilon_0}}
\end{equation}
The field $\phi$ is then given by:
\begin{equation} \phi(x,y)=\tanh\left[ \frac{ g(x,y)}{w}\right]
\label{eq:kinkap}
\end{equation}
where we have absorbed the constant $z_0$ in the function $g$ to
center the wall along the line $(x(s),y(s))$ where $g$ vanishes.
Eq.~(\ref{eq:kinkap}) is  well known  as the field corresponding to a
single wall in the one dimensional $\phi^4$ model.

To calculate the energy of the solution given by
Eq.~(\ref{eq:kinkap}), it is convenient to use as coordinates the
distance $z$ to the center of the interface and $s$, the parameter
defining this center. These new coordinates $(s,z)$  are related
with the cartesian coordinates $(x,y)$ as $(x(s,z),y(s,z))$,
obeying:
\begin{equation} g(x(s,z),y(s,z))=z\quad \forall s,z \label{contc}
\end{equation}
The $z=0$ contour line is our initial curve $(x(s),y(s))$. The
Jacobian of this change of coordinates can be calculated
differentiating Eq.~(\ref{contc}) with respect to $s$ and $z$,
respectively, yielding
\begin{equation} dxdy=\sqrt{\dot x^2+\dot
y^2}ds\,dz
\end{equation}
where the dot denotes differentiation with respect to $s$. This is
in fact the product of the elementary length of the contour line
$\sqrt{\dot x^2+\dot y^2}ds$ times $dz$.

We can now calculate the energy of the wall inserting the solution
(\ref{eq:kinkap}) in Eq.~(\ref{energy0}). With the change of variable
$(x,y)\to (s,z)$, the energy reduces to:
\begin{equation} E=\int dz L(z)
\left[U(f(z))+Hf(z)+\frac{c}{2}f'(z)^2\right]
\end{equation}
where $f(z)=\tanh (z/w)$ and
\begin{equation} L(z)=\int ds \sqrt{\dot x(s,z)^2+\dot y(s,z)^2}
\end{equation}
is the total length of the contour line $g(x,y)=z$ in the restricted
geometry of the problem. In particular, $L(0)$ is the length of our
starting curve $(x(s),y(s))$ defining the center of the interface.
If the interface is narrow, we can approximate $L(z)\simeq L(0)$ for
those $z$ where $U(f(z))$ is significantly different from zero,
i.e., around the center of the interface. Finally, the energy due to
the external field $H$ can be estimated replacing $f(z)$ by a step
function $2\theta(z)-1$ in the term $Hf(z)$. With these assumptions,
the energy becomes:
\begin{equation}
E=\sigma L(0)-2HA
\end{equation}
where $A$ is the area at one side of the center of the interface
$(x(s),y(s))$, and
\begin{eqnarray}
\sigma=\int_{-\infty}^{\infty}\left[U(f(z))+\frac{c}{2}f'(z)^2\right]
=\frac{\sqrt{8\epsilon_0c}}{3}
\end{eqnarray}
is the energy of the interface per unit of length. These are the
expressions yielding \eqref{energialineasinhuecos} in the main text.

\section{Equilibrium shape}\label{sec:basicap}
Our next step is to calculate the equilibrium profile of an
interface segment. The energy of an interface
segment is both a function of  its shape and  the location of its ends
or contact points. In order to find the possible metastable states of the
segment we need to minimize this energy with the constraint
(\ref{perpcondition}).

\begin{figure}
\begin{center}
\includegraphics[angle=0, width=3.5cm]{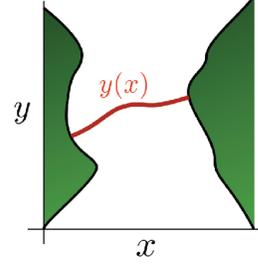}
\caption{(color online) The elastic wall (red online) between two boundaries (vertical black curves) is parametrized as $y(x)$ to solve the variational problem.}
\label{fig:arc2}
\end{center}
\end{figure}

We describe the wall as a line given by $y(x)$, anchored to the
border at points $(x_1,y_1)$ on the left and $(x_2,y_2)$ on the
right (see Fig.~\ref{fig:arc2}). The energy of the interface is given by
\begin{equation}
{E}=\int_{x_1}^{x_2}\left[\sigma\sqrt{1+y'(x)^2}-2Hy(x)\right]\,dx,
\end{equation}
hence the Euler-Lagrange equation is
\begin{equation}
\frac{d}{dx}\frac{ y'}{\sqrt{1+(y')^2}}+\frac{2H}{\sigma} =0,
\label{eq:EL}
\end{equation}
We have to solve this equation imposing the orthogonality condition
at the contact points (which are otherwise free). One integration of
\eqref{eq:EL} gives
\begin{equation}
\frac{y'(x)}{\sqrt{1+y'(x)^2}}=-\frac{x-x_0}{r},
\label{eq:oneint}
\end{equation}
where $x_0$ is a constant and $r=\sigma/(2H)$. From
\eqref{eq:oneint} we get
\begin{equation}
y'(x)=\pm\frac{x-x_0}{\sqrt{r^2-(x-x_0)^2}}
\label{eq:yx}
\end{equation}
and a second integration yields
\begin{equation}
y(x)\pm\sqrt{r^2-(x-x_0)^2}=y_0,
\label{eq:yxy0}
\end{equation}
which, written as $(x-x_0)^2+(y-y_0)^2=r^2$ reveals itself as the
arc of a circumference of radius $r$ and center $(x_0,y_0)$. This is in fact
the Laplace law in two dimensions, relating the pressure
difference to the local curvature of an elastic interface at
equilibrium. It is however important to notice
that this equilibrium shape is independent of the boundary
conditions. We can therefore impose these conditions by looking for an arc of radius $r$ which intersect orthogonally with the two boundaries, as we do in Sec.~\ref{sec:basic} using basic geometric arguments and in Sec.~\ref{sec:depinning} in an analytical manner.

\end{appendix}

\end{document}